\documentclass{article}

\usepackage{amsmath,graphicx,amssymb,fancyhdr,amsthm,enumerate,textcomp,verbatim,cite}
\usepackage[T1]{fontenc}

\def\CH{CH$_0$ }

\def\Journal#1#2#3#4{#4 {\em #1} {\bf #2}, #3}

\def\beq{\begin{eqnarray}}
\def\eeq{\end{eqnarray}}
\def\best{\begin{eqnarray*}}
\def\eest{\end{eqnarray*}}
\def\bcom{}

\def\lc{\overline{\lambda}}

\def\s5{\sqrt{5}}

\def\di{\textrm{d}}

\def\u{{\mathbf u}}
\def\v{{\mathbf v}}

\def\kk{{\mathbf k}}
\def\ll{{\mathbf l}}

\def\P{\Psi}

\def\I{\textrm{i}}

\def\p{\partial}

\def\a{\alpha}

\def\d{\delta}

\def\L{\Lambda}

\def\s{\sigma}

\def\zc{\overline{\zeta}}
\def\zz{\bar{z}}

\def\pu{\p_u}

\def\w{\omega}

\def\Vc{\bar{V}}

\def\mc{\overline{m}}

\def\bar{\overline}

\def\fpd#1#2{\frac{\partial #1}{\partial #2}}
\def\vf#1{\frac{\partial}{\partial{#1}}}

\def\tho{\textrm{\TH}}

\begin{document}

\title{\textbf{Maximally inhomogeneous G\"{o}del-Farnsworth-Kerr generalizations}}


\author{{Lode Wylleman}\\
\vspace{0.05cm} \\
{\small Faculty of Applied Sciences TW16, Ghent
University},  {\small  Galglaan 2, 9000 Gent, Belgium}\\
{\small E-mail: \texttt{lode.wylleman@ugent.be}} }

\date{\today}
\maketitle \pagestyle{fancy}
\fancyhead{} 
\fancyhead[EC]{L. Wylleman} \fancyhead[EL,OR]{\thepage}
\fancyhead[OC]{Inhomogeneous GFK-generalizations}
\fancyfoot{} 

\begin{abstract}
It is pointed out that physically meaningful aligned Petrov type D
perfect fluid space-times with constant zero-order Riemann
invariants are either the homogeneous solutions found by G\"{o}del
(isotropic case) and Farnsworth and Kerr (anisotropic case), or new
inhomogeneous generalizations of these with non-constant rotation.
The construction of the line element and the local geometric
properties for the latter are presented.
\end{abstract}


\section{Introduction}

The results by Milson and Pelavas~\cite{Milson, Milson2} reopened
the question whether the so far theoretically determined Karlhede
upper bounds for given Weyl-Petrov and/or Ricci-Segre type are sharp
as well. In any case, a \emph{necessary} condition is that the
space-time is curvature homogeneous of order zero (further denoted
by \CH), i.e., its zero order Cartan-Riemann invariants are all
constant. It is well known that ample families of pure radiation
metrics satisfy this property~\footnote{I thank Michael Bradley for
reminding me of this after the talk}. Here we revise the situation
for electrovac fields and their Einstein space limits, and present a
theorem classifying all \CH genuine (i.e.\  non-Einstein space)
perfect fluids, in the case where the Weyl tensor is of aligned
Petrov type $D$. We will focus on the physically relevant models,
which turn out to be exhausted by the celebrated homogeneous
G\"{o}del universe, its anisotropic generalizations found by
Farnsworth and Kerr, and a new family constituted by generalizations
of these known solutions, the members of which have
\emph{non-constant} rotation. We mention the construction of good
coordinates, summarize the local properties and end with concrete
and more general conclusions of this investigation.

\section{Inhomogeneous \CH perfect fluid solutions of Petrov type $D$}

In general, \CH Petrov type $D$ space-times are characterized by the
existence of a Weyl principal complex null frame
$(k^a,l^a,m^a,\mc^a)$ relative to which
\begin{equation}\label{Psis}
\P_0=\P_1=\P_3=\P_4=0,\quad \P_2=const\neq 0.
\end{equation}
$k^a$ and $l^a$ spanning the Weyl principal null directions (PND's).

All double aligned Petrov type $D$, non-null Einstein-Maxwell
`electrovac' fields have been classified and integrated by Debever,
Kamran and McLenaghan~\cite{Debeveretal1,Debeveretal2} and
independently by Garcia~\cite{Garcia}. From their results, or from
the invariant approaches in \cite{DebeverMcLen,CzaporMcLen} (see
also \cite{Edgaretal} for a more recent account in the vacuum case,
making use of the GHP formalism) one readily infers that the
\emph{only} \CH solutions in this class are given by
\begin{eqnarray*}
&&\di s^2=\frac{\di x^2}{P(x)}+P(x)\di\phi^2+\frac{\di
y^2}{Q(y)}-Q(y)\di\psi^2,\\
&&P(x)=1-(\L+\Phi_0)x^2,\quad Q(y)=1-(\L-\Phi_0)y^2.
\end{eqnarray*}
They are gravito-electric ($\Psi_2=-\L/3$), homogeneous, have a
complete group $G_6$ of isometries and are attributed to
Levi-Civita~\cite{LeviCiv2}, Robinson~\cite{Robinson} and
Bertotti~\cite{Bertotti}, the last author giving the more general
case with $\L\neq 0$. Putting the electromagnetic field parameter
$\Phi_0$ equal to zero one gets all \CH Einstein spaces (notice that
$\L=0$ gives the Petrov type $O$ Minkowski space-time), having the
same isometry group.

A Petrov type $D$ space-time represents an \emph{aligned perfect
fluid} if its Einstein tensor has the structure
\begin{equation}
G_{ab}=(w+p)u_au_b+p g_{ab},\quad w+p\neq 0\quad u^a=\frac{qk^a+
l^a}{\sqrt{2q}}
\end{equation}
where $q>0$, $w=w'+\L$ and $p=p'+\L$ are the (geometric or
effective) energy density, resp.\ pressure of the fluid (in which
the cosmological constant $\L$ has been absorbed) and $u^a$ is the
fluid 4-velocity, lying in the PND-plane.
Space-times of this nature are \CH if and only if the conditions
(\ref{Psis}) hold and $w$ and $p$ are constant. Then, by
energy-momentum conservation and $w+p\neq 0$, $u^a$ is
non-accelerating and non-expanding. The \emph{only explicitly known}
\CH examples are given by
\begin{eqnarray}\label{GFK}
A^2\di s^2=&&\di u^2-2(\di t+e^x\di y)^2+\frac{e^v}{\cosh v}(\cos\,
t\,\di x+\sin\, t\, e^x\di y)^2\nonumber\\
&&+\frac{e^{-v}}{\cosh v}(-\sin\, t\,\di x +\cos\, t \,e^x\di y)^2
\end{eqnarray}
where $A$ and $v$ are constant. For $v=0$ one obtains the $G_5$
shearfree homogeneous G\"{o}del universe~\cite{Godel}, and for
$v\neq 0$ the anisotropic and shearing, yet still homogeneous
generalizations discovered by Farnsworth and
Kerr~\cite{FarnsworthKerr}. Let us list the local geometric
properties of these space-times:
\begin{itemize}
\item[(A)] the vector field $\v\equiv\frac{q\kk- \ll}{\sqrt{2q}}=\vf{u}$ is covariantly
constant;
\item[(B)] the vorticity and shear of $u^a$ are given by
\begin{eqnarray}\label{wa}
&&\w^a=\w\,\, v^a,\quad \w=a\cosh v\,(=\I
\sqrt{2q}\,\rho=\I\sqrt{{2}/{q}}\,\mu),\\
\label{sab}&&\s_{ab}=V\mc_{(a}\mc_{b)}+\Vc m_{(a}m_{b)},\quad
V=a\sinh
v\,e^{\I\varphi_V}\,(=-\sqrt{2q}\,\sigma=\sqrt{{2}/{q}}\,\lc),
\end{eqnarray}
where the relation with the
NP spin coefficients $\rho$, $\mu$, $\sigma$ and $\lambda$ has been
added, and where $e^{\I\varphi_V}$ is a
spin gauge field;
\item[(C)] the space-time represents gravito-electric `stiff dust', satisfying
\begin{eqnarray}\label{crack}
A^2\equiv p=w=-3\Psi_2=\w_a\w^a-\s_{ab}\s^{ab}=\w^2-V\Vc.
\end{eqnarray}
\end{itemize}
Notice that perfect fluids with $p=w$ satisfy the dominant energy
condition only when $\L\leq 0$ and, when $w$ is moreover constant,
may be interpreted as e.g.\ dust space-times ($\L=-w$) or stiff
fluids ($\L=0$).

The question arises whether the G\"{o}del-Farnsworth-Kerr
space-times are the only \CH aligned Petrov type $D$ perfect fluids.
In contrast to the Einstein space case mentioned above, however,
homogeneity of the curvature invariants does not imply homogeneity
of the space-time here. One can deduce
the following classification result~\cite{WyllGodelgen}:\\

{\bf Theorem.} Any \CH aligned Petrov type $D$ perfect fluid
satisfies property (C). It is either an unphysical member $w=p<0$ of
Ellis' LRS II non-rotating dust family~\cite{EllisLRSdust} or the
Stephani~\cite{Stephani}-Barnes~\cite{Barnes} rotating dust
family, or it satisfies properties (A) and (B) as well.\\

Let us focus on the physically relevant class of space-times
satisfying (A)-(C). When $\w$ is constant one recovers the
homogeneous G\"{o}del-Farnsworth-Kerr solutions; when $\w$ is
non-constant, however, new inhomogeneous solutions arise as follows
(see \cite{WyllGodelgen} for more details). At each point the
variables $\w$, $V$ and $\Vc$ are constrained by the hyperbolic
equation $\w^2-V\Vc=A^2$. Thus they can be parametrized as in
(\ref{wa})-(\ref{sab}), but this does not give suitable coordinates.
A better choice is the parametrization
\begin{eqnarray*}
\w=A\frac{\cosh x}{\cos y}
,\quad V=A\frac{\sin y+\I\sinh x}{\cos y}.
\end{eqnarray*}
Also, the Jacobi identities allow to set the NP spin coefficients
$\a$, $\beta$, $\gamma$ and $\epsilon$ to zero. On rectifying the
vector field $\u=a\,\partial_t$, the integration of the Cartan
equations is quite straightforward. On using $x$, $y$, $t$ and $u$
as coordinates, where $\v=\vf{u}$, one obtains
\begin{eqnarray} A^2\di s^2=&&\di u^2-\left(\di
t+\frac{1}{2}\fpd{F(x,y)}{x}\di
y-\frac{1}{2}\fpd{F(x,y)}{y}\di x\right)^2\nonumber\\
&&+e^{F(x,y)}e^x\left(\cos \left(t-\frac{y}{2}\right)\,\di x+\sin \left(t-\frac{y}{2}\right)\,\di y\right)^2\nonumber\\
&&+e^{F(x,y)}e^{-x}\left(-\sin \left(t+\frac{y}{2}\right)\,\di
x+\cos \left(t+\frac{y}{2}\right)\,\di y\right)^2,\label{metric}
\end{eqnarray}
$F(x,y)$ being a solution of
\begin{equation}
\frac{\partial^2 F(x,y)}{\partial x{}^2}+\frac{\partial^2
F(x,y)}{\partial y{}^2}=4\cosh(x)e^{F(x,y)}.\label{constraint}
\end{equation}
We emphasize that this is the general line element for inhomogeneous
aligned Petrov type $D$ perfect fluids with
positive effective energy density $w$.\\

Finally, denote $t_k$ for the number of functionally independent
components of the Riemann tensor and its first $k$ covariant
derivatives, relative to the canonically fixed frame at step $k$ in
the Karlhede space-time classification algorithm~\cite{Karlhede2},
and let $q$ be the Karlhede bound. As there is at least one Killing
vector field $\vf{u}$ one has $t_q\leq 3$. The \CH assumption means
precisely $t_0=0$, and by direct calculation one finds that $t_1=1$
and $2\leq t_2$, whence $2\leq t_2\leq t_3\leq 3$. Further
investigation then shows that $t_2=t_3$, such that $q=3$ for any
member of the new family. The generic situation is $t_2=3$, in which
case $\vf{u}$ is the only Killing vector field; only a very specific
subfamily satisfies $t_2=2$, in which case the complete isometry
group is abelian $G_2$.

\section{Conclusions}

For physically relevant CH$_0$ aligned Petrov type D perfect fluids
the constant energy density $w$ equals the pressure $p$ on the one
hand, and the difference between the vorticity and shear amplitudes
on the other. In contrast to the homogeneous
G\"{o}del-Farnsworth-Kerr models, these amplitudes can be
non-constant, resulting in the non-constancy of higher-order
curvature invariants and, correspondingly, a dramatic drop of the
isometry group dimension of the relevant 3D part of the metric. A
special choice of (non-invariantly defined) coordinates leads to the
metric (\ref{metric})-(\ref{constraint}).

Put in a much broader context, the results exemplify a
\emph{generation technique}, valid for any metric-based gravitation
theory in any space-time dimension: investigate whether a set of
(possibly higher-order) invariant relations, valid for a well-known
family of space-times, singles out this family. If not, new
interesting families may arise.

\end{document}